\documentclass[cameraready]{Interspeech}

\title{Learnable Classifier-Free Guidance Null Embeddings \\ for Enhanced Controllable Speech Synthesis}

\author[affiliation={1}]{Biel}{Tura-Vecino}
\author[affiliation={1}]{Yoach}{Lacombe}
\author[affiliation={1}]{Julian}{Weber}
\author[affiliation={1}]{Zbigniew}{Łatka}
\author[affiliation={1}]{\\Haitong}{Zhang}
\author[affiliation={1}]{Logan}{Hart}
\author[affiliation={1}]{Eren}{Gölge}

\address{$^1$ Cantina Labs}

\email{biel@cantina.ai, yoach@cantina.ai, eren@cantina.ai}


\usepackage{comment}
\usepackage{graphicx} 
\usepackage{multirow} 
\usepackage{adjustbox}

\begin{document}

\maketitle

\begin{abstract}
Classifier-free Guidance (CFG) is widely adopted in text-to-speech (TTS) systems to enhance generation quality and conditioning fidelity by interpolating between conditioned and unconditioned predictions. A common unconditional technique is to use an empty representation, in the form of a fixed null vector. In this work, we propose replacing this representation with a learnable unconditional embedding, optimized to represent a meaningful unconditional state. Objective and subjective evaluations demonstrate that learnable null embeddings consistently outperform fixed null embeddings across speaker similarity, speech stability, and expressiveness, while exhibiting greater robustness to larger guidance scales. We further show that learning a distinct unconditional embedding for each of the TTS conditioning modalities allows fine-grained control over speaker and text guidance, showcasing the trade-off between similarity and quality, and stability and expressiveness in the generated speech. \footnote{Blogpost article with TTS samples available in \url{https://airtimemedia.github.io/IS2026-LearnableCFG/}.} 
\end{abstract}

\section{Introduction}

Current large-scale text-to-speech (TTS) models rely on auto-regressive (AR) modeling of speech representations through a Large Language Model (LLM) backbone \cite{valle, llasa}, optionally coupled with an audio refinement module \cite{audiolm}, which can take the form of transformer-based heads \cite{rq_transformer, maskgct}, lightweight diffusion heads \cite{ditar, vibevoice, voxcpm2}, or full conditional flow-matching decoders \cite{cosy, indextts25}. Regardless of the architecture, the two minimal and most commonly used conditioning signals are the text to be synthesized and the target voice, or speaker identity, in which the speech should be generated. Given its conditional generative nature, TTS has been shown to benefit significantly from the application of Classifier-free Guidance (CFG) \cite{koel_tts, dualspeech, megatts3}, a technique widely adopted across generative modeling domains to enhance output quality and conditioning fidelity \cite{stay_cfg}.

CFG operates by dynamically interpolating between conditioned and unconditioned model outputs at inference time, effectively steering generation toward the desired conditioning signals. Reliable use of CFG requires the model to be robust to unconditional generation at inference time,  which in turn necessitates some form of conditioning dropout during training \cite{cfg}. There are two practical ways of implementing this in auto-regressive TTS: 1) dropping the conditioning on the whole training batch and perform two independent inference passes (conditional and unconditional) to combine the predictions \cite{f5tts, voiceldm}, or 2) mask the conditionings at training time and perform a single 2-batched inference with custom attention masks \cite{ditar}. The latter is generally preferred, as it leverages batched inference and parallel computation \cite{spotify}. However, custom dynamic attention masks that do not follow a fixed bidirectional or fully causal pattern leads to suboptimal performance with compiled inference frameworks, as they may trigger repeated recompilations or prevent backbone optimizations \cite{flashattention}.

To mitigate this issue, the masking approach can be replaced by substituting the conditions with a fixed representation \cite{ditar}, known as the unconditional vector/token, typically initialized to zeros \cite{voxcpm2}. While this approach preserves the causal attention pattern and avoids compilation overhead, it presents two significant limitations in TTS with multiple conditionings. First, using a single unconditional vector fails to distinguish between fundamentally different conditioning types: in TTS, speaker identity and linguistic content are orthogonal signals that independently control distinct aspects of the generated speech. Second, a predefined unconditional vector may lie outside the model's input training distribution, potentially introducing training instabilities through large gradients or, in the case of zero vectors, numerical instabilities at inference time.


In this work, we propose replacing the fixed unconditional vector with a set of learnable null embeddings, one per con- ditioning signal, each representing the absence of that specific condition. Unlike fixed vectors, learnable null embeddings nat- urally adapt to the model’s training distribution, converging toward a stable and meaningful unconditional baseline within their own conditioning domain. This also enables implicit gra- dient sharing across conditioning axes: when the text condi- tioning is replaced by its learnable null embedding, the latter still receives gradients from the speaker-only conditional state, and vice versa, ensuring that each null embedding is optimized by the model’s partially conditioned end-to-end training signal. 

Beyond improving CFG-based generation, learnable null embeddings enable fine-grained attribute control over the generated speech. By independently manipulating the CFG strength for each conditioning modality \cite{voiceldm}, we show that learnable null embeddings provide a robust method for disentangling the effects of each conditioning attribute on the generated speech.

Therefore, the main contributions of this work are:
\begin{itemize}
\item A learnable CFG null embedding technique that provides a more stable and meaningful unconditional state, improving speaker similarity, speech stability, and expressiveness over the fixed null embedding baseline.
\item A decoupled CFG formulation that independently controls the guidance strength of each condition, enabling fine-grained attribute control over different aspects of the generated speech and revealing generation trade-offs between similarity/quality and stability/expressiveness.
\end{itemize}

\newpage

\section{Methodology}

\subsection{Model}
\label{sec:model}
Our TTS model consists of two main components: 1) an AR GPT Qwen3-based 0.6B backbone \cite{qwen3} with lightweight diffusion heads, following the next-token diffusion paradigm \cite{ditar, voxcpm2, kyutai_calm}, and 2) a causal transformer-based variational autoencoder (VAE) \cite{vibevoice} that encodes speech into low-dimensional 64-dim latent vectors $z$ and decodes them back to high-quality audio. The GPT backbone, $g_\theta$, is conditioned on speaker latents $s$, obtained by encoding a reference mel-spectrogram through a Perceiver encoder \cite{xtts}, and on BPE-compressed text tokens $t$ representing the text to be synthesized \cite{bpe}. At each AR step $i$, the GPT acoustic head is trained to predict a binary modes: generate or stop, framed as a binary cross-entropy classification problem. When in generation mode, the last hidden state $h_i$ is passed as conditioning to a diffusion head parameterized by a lightweight MLP $\epsilon_\theta$, which predicts the target VAE latent $z_i$ for that frame through iterative denoising. This latent is then normalized to mitigate AR error accumulation \cite{hyperspherical} and fed back into $g_\theta$ to continue generation until the stop token is predicted.

\subsection{Classifier-Free Guidance (CFG)}
\label{sec:cfg}
At inference time, the quality and adherence to the conditioning signals can be enhanced via CFG. We adopt a conditional baseline CFG formulation in the diffusion heads, which steers the prediction away from an unconditional reference \cite{ltx2}:
\begin{equation}
    \hat{\epsilon}_\theta = \epsilon_\theta(z_i, h_i) + w \left(\epsilon_\theta(z_i, h_i) - \epsilon_\theta(z_i, \bar{h}_i)\right)
    \label{eq:cfg}
\end{equation}
where $h_i$ and $\bar{h}_i$ denote the conditional and unconditional representations from the backbone GPT hidden states, defined as:
\begin{equation}
    h_i = g_\theta(z_{:i-1}, s, t) \hspace{2mm} \text{,} \hspace{2mm} \bar{h}_i = g_\theta(z_{:i-1}, \emptyset)
\end{equation}
and $w \geq 0$ controls the degree of steering away from the unconditional prediction. This formulation ensures that the conditional estimate serves as the baseline, with the unconditional prediction used as a reference to amplify the condition effect.

\subsubsection{Learnable Null Embedding}
In standard CFG, the unconditional hidden state $\bar{h}_i$ is obtained by dropping all conditioning signals, typically replacing them with a fixed null embedding $\emptyset$. During training, the model is made robust to null conditioning by dropping the conditioning signals with a certain probability \cite{cfg}.
Instead, we propose to replace this fixed unconditional representation with a learnable null embedding for each conditioning modality: $\bar{s}$ for the speaker and $\bar{t}$ for the text. Thus, the unconditional state follows:
\begin{equation}
    \bar{h}_i = g_\theta(z_{:i-1}, \bar{s}, \bar{t})
\end{equation}
During training, these null embeddings are optimized independently by replacing the conditioning signals with a fixed probability, allowing the model to learn a stable in-domain unconditional baseline compared to a fixed, pre-defined representation.

\subsubsection{Independent Attribute CFG}
Since each conditioning modality represents a distinct speech attribute, their guidance contributions can be decoupled and controlled independently. Similar to dual-CFG notation \cite{dualspeech}, we define modality-specific unconditional hidden states as:
\begin{equation}
    \bar{h}^s_i = g_\theta(z_{:i-1}, \bar{s}, t) \hspace{2mm} \text{,} \hspace{2mm} \bar{h}^t_i = g_\theta(z_{:i-1}, s, \bar{t})
\end{equation}
This allows us to extend Equation \eqref{eq:cfg} to apply decoupled guidance for each attribute independently:
\begin{equation}
\begin{split}
    \hat{\epsilon}_\theta = \epsilon_\theta(z_i, h_i) 
    &+ w_{s} \left(\epsilon_\theta(z_i, h_i) - \epsilon_\theta(z_i, \bar{h}^s_i)\right) \\
    &+ w_{t} \left(\epsilon_\theta(z_i, h_i) - \epsilon_\theta(z_i, \bar{h}^t_i)\right)
\end{split}
    \label{eq:cfg_2}
\end{equation}
where $w_s \geq 0$ and $w_t \geq 0$ independently control the strength of speaker and text guidance, enabling fine-grained control over each conditioning strength separately.


\section{Experiments and results}

\subsection{Experimental setup}

We train two variants of the same TTS model described in Section \ref{sec:model} for the same number of steps and on the same seeded data subset, using a combination of a binary cross-entropy loss, predicting whether to generate speech or stop, on the AR backbone acoustic head and an end-to-end MSE loss on the diffusion heads that back-propagates through the full model. To ensure robustness to CFG, we drop each condition independently with a probability of $0.1$ during training, following the minimal optimal dropout value in \cite{cfg}. In the first variant, dropped conditionings are replaced with a fixed zero vector representing null conditioning. In the second, each conditioning is replaced with a dedicated learnable embedding, initialized with a standard normal distribution and jointly optimized with the model.

For the evaluation, we curated 65 unseen, expressive, proprietary characterful speaker references and synthesized 2 new sentences per speaker: 130 generated samples\footnote{We use custom evaluations instead of open benchmarks \cite{seedtts}, which suffer from low quality, monotonous styles, and training data overlap risks. More details on our evaluation suite are left to future work.}. To contextualize the reported metrics, we compare our model against several open-source SOTA TTS systems, both without CFG: \textit{Moss-TTS Local} \cite{mosstts} and \textit{Qwen3-TTS 0.6B} \cite{qwen3tts}, and with CFG: \textit{VibeVoice 1.5B} \cite{vibevoice} \textit{VoxCPM v2} \cite{voxcpm2}, \textit{dots.tts} \cite{dotstts} and \textit{IndexTTS v2.5} \cite{indextts25}.

\subsection{Evaluation metrics}

For objective metrics, we measure: 1) \textit{Speech stability}, reported as the Character Error Rate (CER) between the target text and the transcription obtained through Whisper v3-large \cite{whisper}. 2) \textit{Speaker similarity}, computed as the cosine distance between reference and generated speaker embeddings using the pre-trained ECAPA2 speaker embedding model \cite{ecapa2} (SECS) and a pre-trained WavLM-based prosody embedding model \cite{wavlm_tbr_pro} (PRO), along with the Pitch Mean Ratio (PMR) between reference and generated speech \cite{fcpe_pitch}. 3) \textit{Speech quality}, for which we report predicted Production Quality of the speech (PQ) \cite{audiobox} and the predicted Mean Opinion Score (UTMOS) \cite{utmos}. 4) \textit{Speech dynamics}, reported as the Pitch Standard Deviation of the generated speech (Pitch Std), and the Speech Rate Ratio (SRR), computed as the ratio of the number of transcribed words per unit of non-silence parts between generated and reference speech.

For the subjective evaluation, we assessed speech naturalness and speaker similarity to a reference audio using a Comparative Mean Opinion Score (CMOS) test in a multi-model setup. Three CFG model variants were compared pairwise: fixed zero embedding and two learnable null embedding: static and tuned guidance. For each pairwise comparison, 90 annotators evaluated 20 evaluation cases, each consisting of two samples, A and B, generated by different systems. Annotators were asked to rate their relative preference in terms of naturalness and speaker similarity on a scale from –2 (Sample A) to +2 (Sample B) \cite{prolific}.

\begin{figure*}[t]
  \centering
  \begin{adjustbox}{center}
    \includegraphics[width=1.05\textwidth]{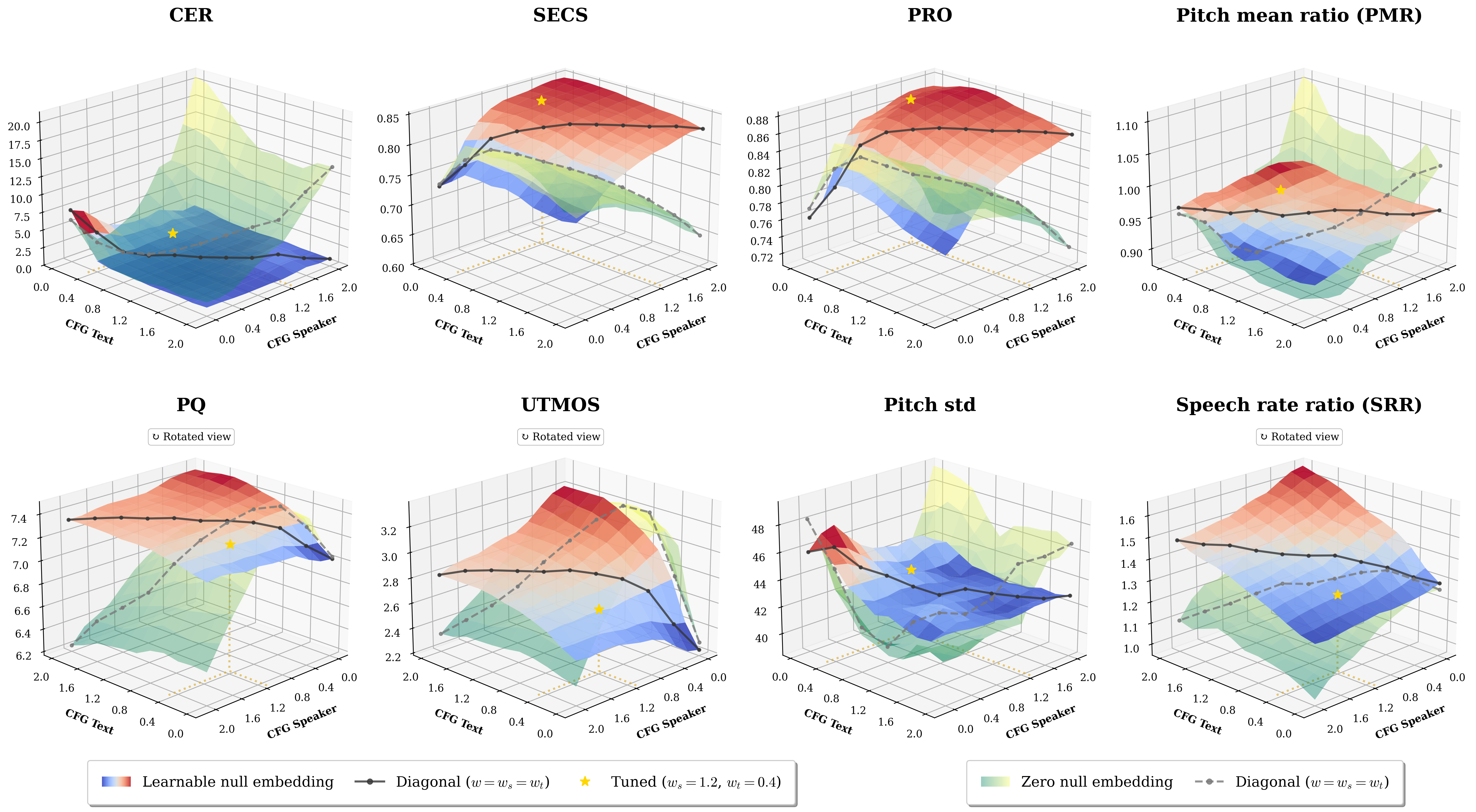}
  \end{adjustbox}
  \caption{3D heatmaps illustrating the objective metric landscape as a function of speaker and text CFG weights, evaluated using learnable and fixed zero null embeddings. The diagonal curve traces the manifold corresponding to standard (non-disentangled) CFG, where speaker and text guidance are coupled and varied jointly ($w = w_s = w_t$). The yellow star marks the custom hand picked decoupled values of CFG guidance that we report in our evaluations. Note that some axes are rotated for better visualization.}
  \label{fig:heatmaps}
\end{figure*}

\newpage

\begin{table*}[th]
  \caption{Results from objective metrics comparing state-of-the-art open-source TTS models and our proposed architecture, illustrating the impact of different CFG methods. Underlined metrics indicate the overall best performance, while bolded metrics represent the best performance across our model ablations. Quality metrics include a correlation coefficient (in brackets) with respect to the reference.}
  \label{tab:objective_metrics}
  \centering
  \resizebox{\linewidth}{!}{%
  \begin{tabular}{l l | c c c c c c c c}
    \toprule
    \textbf{Model configuration} & \textbf{CFG guidance} & \textbf{CER $\downarrow$} & \textbf{SECS $\uparrow$} & \textbf{PRO $\uparrow$} & \textbf{PMR $\sim1$} & \textbf{PQ $\uparrow$} & \textbf{UTMOS $\uparrow$} & \textbf{Pitch std $\uparrow$} & \textbf{SRR $\sim1$} \\
    \midrule
    Qwen3-TTS 0.6B \cite{qwen3tts} & $-$         & $6.9$ & $0.743$ & $0.814$ & $0.95$ & $\underline{7.74} \hspace{1mm} (0.60)$ & $\underline{3.50} \hspace{1mm} (0.71)$ & $45.1$ & $1.38$  \\
    Moss-TTS \cite{mosstts} & $-$         & $1.1$ & $0.762$ & $0.797$ & $0.91$ & $7.47 \hspace{1mm} (0.65)$ & $3.26 \hspace{1mm} (0.78)$ & $41.3$ & $1.24$  \\
    VibeVoice \cite{vibevoice} & $w = 1.3$         & $3.3$ & $0.701$ & $0.772$ & $0.93$ & $7.55 \hspace{1mm} (0.62)$ & $3.11 \hspace{1mm} (0.79)$ & $44.4$ & $1.22$  \\
    VoxCPM v2 \cite{voxcpm2} & $w = 1.0$         & $1.4$ & $0.776$ & $0.820$ & $1.01$ & $7.14 \hspace{1mm} (0.64)$ & $2.85 \hspace{1mm} (0.80)$ & $41.1$ & $1.21$  \\
    dots.tts \cite{dotstts} & $w = 1.2$ & $3.2$ & $0.815$ & $0.825$ & $1.05$ & $7.40 \hspace{1mm} (0.69)$ & $2.86 \hspace{1mm} (0.81)$ & $45.4$ & $\underline{1.15}$  \\
    IndexTTS v2.5 \cite{indextts25} & $w = 0.7$      & $2.2$ & $0.763$ & $0.692$ & $\underline{1.00}$ & $7.12 \hspace{1mm} (0.68)$ & $2.63 \hspace{1mm} (\underline{0.95})$ & $44.1$ & $1.17$  \\
    
    \midrule
    Baseline (w/o CFG) & $w = 0$                        & $7.3$ & $0.725$ & $0.758$ & $0.94$ & $7.02 \hspace{1mm} (0.67)$ & $2.23 \hspace{1mm} (0.82)$ & $\underline{\mathbf{45.8}}$ & $\mathbf{1.30}$  \\
    Fixed Zero embed. & $w = 0.8$                       & $1.2$ & $0.755$ & $0.807$ & $0.88$ & $7.20 \hspace{1mm} (0.66)$ & $\mathbf{3.22} \hspace{1mm} (0.76)$ & $37.9$ & $1.34$  \\
    Learnable Null embed. & $w = 0.8$                   & $\underline{\mathbf{0.9}}$ & $0.817$ & $0.862$ & $0.95$ & $\mathbf{7.33} \hspace{1mm} (\underline{\mathbf{0.79}})$ & $2.82 \hspace{1mm} (0.93)$ & $42.6$ & $1.41$  \\
    Learnable Null embed. & $w_{t} = 0.4, w_{s} = 1.2$  & $1.2$ & $\underline{\mathbf{0.841}}$ & $\underline{\mathbf{0.877}}$ & $\mathbf{0.96}$ & $7.27 \hspace{1mm} (0.77)$ & $2.75 \hspace{1mm} (\mathbf{0.94})$ & $43.5$ & $1.33$  \\
    \bottomrule
    \end{tabular}}
\end{table*}

\subsection{Fixed versus learnable null embedding}

To assess whether learnable null embeddings yield a consistent improvement in generation quality, we first consider the coupled CFG setting, where a single shared guidance weight $w$ is applied to all conditioning signals. The effect of varying $w$ is illustrated in Figure \ref{fig:heatmaps}, where the diagonal cross-sections plot of the 3D surfaces correspond to this coupled trajectory. As shown in the line plots, the fixed zero embedding is more sensitive to the guidance scale compared to the learnable null embedding, with generation quality degrading sharply for $w > 0.8$, manifesting as reduced perceptual quality (PQ and UTMOS) and decreased speaker similarity. For small guidance values, the fixed zero embedding does improve across all metrics, confirming that traditional fixed null conditioning CFG scheme enhances TTS generations. In contrast, the learnable null embedding exhibits similar behavior at low guidance strengths but reaches a stable quality plateau that is substantially more robust to larger values of $w \geq 1.0$. Notably, for speaker similarity metrics (SECS, PRO\_SECS), this plateau consistently exceeds the peak performance of the fixed zero embedding, suggesting that a learned unconditional baseline provides a stronger and more stable reference from which to steer the conditional guidance.

To further quantify these differences, we select the guidance value that maximizes all objective metrics for the zero-embedding CFG baseline, $w=0.8$, and compare it against its learnable null embedding counterpart in Table \ref{tab:objective_metrics}. In terms of speaker similarity, the learnable null embedding consistently outperforms the fixed zero embedding across all three metrics: SECS, PRO, and Pitch Mean Ratio (PMR). Regarding speech stability, both models achieve comparably low CER, with no statistically meaningful difference between them. Notably, the learnable null embedding exhibits higher pitch variations (Pitch std) at the same CER level, indicating that the generated speech is more expressive without compromising intelligibility. In terms of perceptual quality, the fixed zero embedding reports lower PQ and higher UTMOS scores; these metrics reflect absolute signal quality rather than faithfulness to the reference speaker and should be interpreted with caution. To account for this, we additionally report quality correlation metrics with respect to the reference speaker, which provide a more direct measure of conditioning adherence. These reveal that the learnable null embedding, while producing lower absolute quality scores, generates more faithful speech. This observation is supported by both superior similarity metrics and higher quality correlations with respect to the reference. 

These findings are further supported by the subjective CMOS results reported in Table \ref{tab:subjective_metrics}. Both variants of the learnable null embedding obtain positive CMOS scores for naturalness and speaker similarity, indicating a consistent listener preference over the fixed zero embedding baseline, which is the least favored model and reports negative CMOS values for both.

\subsection{Speech attribute control analysis} 
As described in Section \ref{sec:cfg}, the effect of each conditioning signal on the generated speech can be explored by increasing the inference batch size to 3 and independently tuning the speaker, $w_s$, and text, $w_t$, guidance weights. Beyond the coupled diagonal ($w = w_s = w_t$), Figure \ref{fig:heatmaps} illustrates the full objective metric manifold obtained by combining different guidance scales independently, revealing the behavior of each metric across the joint guidance space. Comparing the two surfaces per objective metric, we confirm that the learnable null embedding produces smoother transitions and scales in a more robust manner to larger guidance values compared to the fixed zero embedding, where both CER and SECS diverge for larger CFG weights. As the learnable null embedding is a stronger unconditional baseline, we focus on this variant for our attribute control analysis.

For the text guidance axis, CER decreases rapidly as $w_t$ increases. Interestingly, pitch standard deviation follows an inverse correlation with CER along the same axis: more expressive speech tends to exhibit higher transcription error rates, likely due to increased mispronunciation or greater deviation from monotonic intonations. This reveals a first clear trade-off between stability and expressiveness, primarily driven by the text CFG weight. Additionally, lower values of $w_t$ tend to yield lower perceptual quality scores, while higher values produce more conservative speech with reduced prosodic variation but higher predicted quality metrics.

For speaker similarity, it is $w_s$ that drives SECS, PRO, and Pitch Mean Ratio (PMR), with all three metrics improving as speaker guidance increases. Higher similarity influences the quality correlation metrics, as the generated speech aligns more closely with the characteristics of the reference. Consequently, PQ and UTMOS are partially modulated by $w_s$, showing the interplay between speaker fidelity and speech quality.

A similar coupling is observed for the Speech Rate Ratio (SRR): stronger text CFG directly increases the speech rate of the generated output \cite{parakeet}, and this effect is further amplified by higher values of $w_s$ at large $w_t$.

Empirically, we find that a high speaker guidance weight combined with a moderate text guidance weight provides a favorable trade-off for faithful and expressive speech generation respectively. Based on this, we tuned $w_s=1.2$ and $w_t=0.4$ and report the corresponding objective metrics in Table \ref{tab:objective_metrics}, as well as illustrating, in Figure \ref{fig:heatmaps}, this combination as a single star mark. This configuration yields consistent objective improvements compared to the coupled single weight guidance $w$ variant. Specifically, for this combination, the model produces speech that is faithful to the reference speaker, maintains a low CER while exhibiting strong pitch variation, and achieves a more controlled speech rate. All of these improvements come at the cost of a marginal reduction in absolute perceptual quality scores, which, as discussed, reflects better adherence to the reference, as evidenced by the quality correlation metrics, rather than a degradation in generation capability.

We included both learnable embedding variants in the multi-model subjective listening tests. While both outperform the fixed zero embedding baseline, a perceptual trade-off is shown between them in Table \ref{tab:subjective_metrics}. The single CFG variant ($w$) achieves higher perceived naturalness, whereas the tuned ($w_s$, $w_t$) variant shows a clearly stronger speaker similarity preference. Based on informal listening, we attribute this discrepancy to two main factors: 1) the tuned variant reports slightly lower absolute quality metrics than the coupled $w$ model, which could influence naturalness judgments, as cleaner signals are often perceived as more pleasant, and 2) given that our test set primarily consists of expressive and characterful speakers, accurately following their prosodic variability, reflected in increased pitch variations and higher similarity, may occasionally be perceived as less natural compared to more conservative, were speech is more intelligible but less speaker-faithful generations.

\begin{table}[th]
  \caption{Results of the subjective multi-model pair-wise CMOS evaluation. Scores are reported as mean ± 95\% confidence interval. Bold indicates the overall preferred model.}
  \label{tab:subjective_metrics}
  \centering
  \resizebox{\linewidth}{!}{%
  \begin{tabular}{l | r r}
    \toprule
    \textbf{Model configuration} & \textbf{Naturalness} & \textbf{Similarity} \\
    \midrule
    Fixed Zero embed. ($w$) &  $-0.171 \pm 0.03$ & $-0.318 \pm 0.03$ \\
    Learnable Null embed. ($w$) & $\mathbf{0.130} \pm 0.02$ & $0.081 \pm 0.03$ \\
    Learnable Null embed. ($w_s$, $w_t$) & $0.106 \pm 0.03$ & $\mathbf{0.178} \pm 0.03$ \\
    \bottomrule
    \end{tabular}}
\end{table}

\section{Conclusions}

We presented a simple yet effective technique that enhances CFG in TTS models by replacing the fixed unconditional representation with a learnable null embedding, providing a more meaningful unconditional baseline from which to steer the guidance. We demonstrated that, under the same experimental setup, the proposed approach surpasses its fixed counterpart in both objective metrics and subjective listening tests while being more robust to larger guidance values. We further demonstrated that, by introducing two independent learnable unconditional states, specific attributes of the generated speech can be controlled by tuning the guidance of each conditioning signal. We showed that text guidance exhibits a trade-off between speech stability and expressivity, while speaker guidance presents a similarity-quality trade-off, making the guidance scale an inference-time hyperparameter for controlling distinct attributes. Tuning these scales leads to a clear improvement in objective metrics, while subjective tests reveal the trade-off between similarity and naturalness when adjusting them independently.

\newpage


\bibliographystyle{IEEEtran}
\bibliography{mybib}

@inproceedings{ecapa2,
    title={ECAPA2: A hybrid neural network architecture and training strategy for robust speaker embeddings},
    author={Thienpondt, Jenthe and Demuynck, Kris},
    booktitle={Automatic speech recognition and understanding (ASRU)},
    year={2023},
}

@inproceedings{whisper,
  title={Robust speech recognition via large-scale weak supervision},
  author={Radford, Alec and Kim, Jong Wook and Xu, Tao and Brockman, Greg and McLeavey, Christine and Sutskever, Ilya},
  booktitle={International conference on machine learning (ICML)},
  year={2023},
}

@inproceedings{wavlm_tbr_pro,
  title={Disentangling prosody and timbre embeddings via voice conversion.},
  author={Gengembre, Nicolas and Le Blouch, Olivier and Gendrot, Cedric},
  booktitle={Interspeech},
  year={2024}
}

@misc{audiobox,
  title={Meta Audiobox Aesthetics: Unified Automatic Quality Assessment for Speech, Music, and Sound}, 
  author={Andros Tjandra and Yi-Chiao Wu and Baishan Guo and John Hoffman and Brian Ellis and Apoorv Vyas and Bowen Shi and Sanyuan Chen and Matt Le and Nick Zacharov and Carleigh Wood and Ann Lee and Wei-Ning Hsu},
  year={2025},
  eprint={2502.05139},
  archivePrefix={arXiv},
}

@inproceedings{utmos,
  title={Utmos: Utokyo-sarulab system for voicemos challenge 2022},
  author={Saeki, Takaaki and Xin, Detai and Nakata, Wataru and Koriyama, Tomoki and Takamichi, Shinnosuke and Saruwatari, Hiroshi},
  booktitle={Interspeech},
  year={2022}
}

@inproceedings{xtts,
  title={XTTS: a massively multilingual zero-shot text-to-speech model},
  author={Casanova, Edresson and Davis, Kelly and G{\"o}lge, Eren and G{\"o}knar, G{\"o}rkem and Gulea, Iulian and Hart, Logan and Aljafari, Aya and Meyer, Joshua and Morais, Reuben and Olayemi, Samuel and others},
  booktitle={Interspeech},
  year={2024}
}

@inproceedings{ditar,
  title={Ditar: Diffusion transformer autoregressive modeling for speech generation},
  author={Jia, Dongya and Chen, Zhuo and Chen, Jiawei and Du, Chenpeng and Wu, Jian and Cong, Jian and Zhuang, Xiaobin and Li, Chumin and Wei, Zhen and Wang, Yuping and others},
  booktitle={International conference on machine learning (ICML)},
  year={2025}
}

@misc{vibevoice,
      title={VibeVoice Technical Report}, 
      author={Zhiliang Peng and Jianwei Yu and Wenhui Wang and Yaoyao Chang and Yutao Sun and Li Dong and Yi Zhu and Weijiang Xu and Hangbo Bao and Zehua Wang and Shaohan Huang and Yan Xia and Furu Wei},
      year={2025},
      eprint={2508.19205},
      archivePrefix={arXiv},
      primaryClass={cs.CL},
      url={https://arxiv.org/abs/2508.19205}, 
}

@inproceedings{kyutai_calm,
  title={Continuous audio language models},
  author={Rouard, Simon and Orsini, Manu and Roebel, Axel and Zeghidour, Neil and D{\'e}fossez, Alexandre},
  booktitle={International conference on learning representations (ICLR)},
  year={2025}
}

@inproceedings{dualspeech,
  title={Dualspeech: Enhancing speaker-fidelity and text-intelligibility through dual classifier-free guidance},
  author={Yang, Jinhyeok and Lee, Junhyeok and Choi, Hyeong-Seok and Ji, Seunghun and Kim, Hyeongju and Lee, Juheon},
  booktitle={Interspeech},
  year={2024}
}

@misc{megatts3,
  title={Megatts 3: Sparse alignment enhanced latent diffusion transformer for zero-shot speech synthesis},
  author={Jiang, Ziyue and Ren, Yi and Li, Ruiqi and Ji, Shengpeng and Zhang, Boyang and Ye, Zhenhui and Zhang, Chen and Jionghao, Bai and Yang, Xiaoda and Zuo, Jialong and others},
  journal={arXiv preprint arXiv:2502.18924},
  year={2025}
}

@misc{ltx2,
  title={LTX-2: Efficient joint audio-visual foundation model},
  author={HaCohen, Yoav and Brazowski, Benny and Chiprut, Nisan and Bitterman, Yaki and Kvochko, Andrew and Berkowitz, Avishai and Shalem, Daniel and Lifschitz, Daphna and Moshe, Dudu and Porat, Eitan and others},
  journal={URL https://arxiv. org/abs/2601.03233},
  year={2026}
}

@inproceedings{spotify,
  title={Knowledge distillation for Transformer-based text-to-speech models},
  author={Henriksson, Erik and Merritt, Thomas and Dall, Rasmus and Vaughan, Felix and Morfi, Veronica},
  booktitle={Speech Synthesis Workshop (SSW)},
  year={2025}
}

@article{cfg,
  title={Classifier-free diffusion guidance},
  author={Ho, Jonathan and Salimans, Tim},
  journal={Conference on Neural Information Processing Systems (NeurIPS)},
  year={2022}
}

@article{koel_tts,
  title={Koel-TTS: Enhancing LLM based Speech Generation with Preference Alignment and Classifier Free Guidance}, 
  author={Shehzeen Hussain and Paarth Neekhara and Xuesong Yang and Edresson Casanova and Subhankar Ghosh and Mikyas T. Desta and Roy Fejgin and Rafael Valle and Jason Li},
  journal={Workshop on Machine Learning for Audio (ICML)},
  year={2025},
}

@article{stay_cfg,
  title={Stay on topic with classifier-free guidance},
  author={Sanchez, Guillaume and Fan, Honglu and Spangher, Alexander and Levi, Elad and Ammanamanchi, Pawan Sasanka and Biderman, Stella},
  journal={International Conference on Machine Learning (ICML)},
  year={2024}
}

@article{fcpe_pitch,
  title={FCPE: A Fast Context-based Pitch Estimation Model},
  author={Luo, Yuxin and Zhang, Ruoyi and Liu, Lu-Chuan and Li, Tianyu and Liu, Hangyu},
  journal={arXiv preprint arXiv:2509.15140},
  year={2025}
}

@article{cosy,
  title={Cosyvoice 3: Towards in-the-wild speech generation via scaling-up and post-training},
  author={Du, Zhihao and Gao, Changfeng and Wang, Yuxuan and Yu, Fan and Zhao, Tianyu and Wang, Hao and Lv, Xiang and Wang, Hui and Ni, Chongjia and Shi, Xian and others},
  journal={arXiv preprint arXiv:2505.17589},
  year={2025}
}

@article{indextts25,
  title={Indextts 2.5 technical report},
  author={Li, Yunpei and Zhou, Xun and Wang, Jinchao and Wang, Lu and Wu, Yong and Zhou, Siyi and Zhou, Yiquan and Wang, Yining and Yang, Yaogen and Hu, Zhetao and others},
  journal={arXiv preprint arXiv:2601.03888},
  year={2026}
}

@article{dotstts,
  title={dots. tts Technical Report},
  author={Lian, Shi and Li, Changtao and Li, Bohan and Wang, Hankun and Zheng, Da and Tian, Junfeng and Ma, Yufeng and Zhang, Colin and Yu, Kai},
  journal={arXiv preprint arXiv:2606.07080},
  year={2026}
}

@article{voxcpm2,
  title={Voxcpm2 technical report},
  author={Zhou, Yixuan and Zeng, Guoyang and Liu, Xin and Li, Xiang and Yu, Renjie and Gui, Jiancheng and Wu, Jiaheng and Wang, Ziyang and Shen, Xudong and Ye, Runchuan and others},
  journal={arXiv preprint arXiv:2606.06928},
  year={2026}
}

@article{mosstts,
  title={Moss-tts technical report},
  author={Gong, Yitian and Jiang, Botian and Zhao, Yiwei and Yuan, Yucheng and Chen, Kuangwei and Jiang, Yaozhou and Chang, Cheng and Hong, Dong and Chen, Mingshu and Li, Ruixiao and others},
  journal={arXiv preprint arXiv:2603.18090},
  year={2026}
}

@article{rq_transformer,
  title={Quantize more, lose less: Autoregressive generation from residually quantized speech representations},
  author={Han, Yichen and Hao, Xiaoyang and Chen, Keming and Xiong, Weibo and He, Jun and Zhang, Ruonan and Cao, Junjie and Liu, Yue and Li, Bowen and Zhang, Dongrui and others},
  journal={arXiv preprint arXiv:2507.12197},
  year={2025}
}

@article{valle,
  title={Neural codec language models are zero-shot text to speech synthesizers},
  author={Wang, Chengyi and Chen, Sanyuan and Wu, Yu and Zhang, Ziqiang and Zhou, Long and Liu, Shujie and Chen, Zhuo and Liu, Yanqing and Wang, Huaming and Li, Jinyu and others},
  journal={arXiv preprint arXiv:2301.02111},
  year={2023}
}

@article{llasa,
  title={Llasa: Scaling train-time and inference-time compute for llama-based speech synthesis},
  author={Ye, Zhen and Zhu, Xinfa and Chan, Chi-Min and Wang, Xinsheng and Tan, Xu and Lei, Jiahe and Peng, Yi and Liu, Haohe and Jin, Yizhu and Dai, Zheqi and others},
  journal={arXiv preprint arXiv:2502.04128},
  year={2025}
}

@article{audiolm,
  title={Audiolm: a language modeling approach to audio generation},
  author={Borsos, Zal{\'a}n and Marinier, Rapha{\"e}l and Vincent, Damien and Kharitonov, Eugene and Pietquin, Olivier and Sharifi, Matt and Roblek, Dominik and Teboul, Olivier and Grangier, David and Tagliasacchi, Marco and others},
  journal={IEEE/ACM transactions on audio, speech, and language processing},
  year={2023}
}

@misc{maskgct,
  title={MaskGCT: Zero-Shot Text-to-Speech with Masked Generative Codec Transformer}, 
  author={Yuancheng Wang and Haoyue Zhan and Liwei Liu and Ruihong Zeng and Haotian Guo and Jiachen Zheng and Qiang Zhang and Xueyao Zhang and Shunsi Zhang and Zhizheng Wu},
  year={2024},
  url={https://arxiv.org/abs/2409.00750}, 
}

@misc{parakeet,
  title={Parakeet},
  author={Darefsky, Jordan and Zhu, Ge and Duan, Zhiyao},
  year = {2024},
  url = {https://jordandarefsky.com/blog/2024/parakeet/}
}

@inproceedings{f5tts,
  title={F5-tts: A fairytaler that fakes fluent and faithful speech with flow matching},
  author={Chen, Yushen and Niu, Zhikang and Ma, Ziyang and Deng, Keqi and Wang, Chunhui and JianZhao, JianZhao and Yu, Kai and Chen, Xie},
  booktitle={Annual Meeting of the Association for Computational Linguistics (ACL)},
  year={2025}
}

@inproceedings{voiceldm,
  title={Voiceldm: Text-to-speech with environmental context},
  author={Lee, Yeonghyeon and Yeon, Inmo and Nam, Juhan and Chung, Joon Son},
  booktitle={International Conference on Acoustics, Speech and Signal Processing (ICASSP)},
  pages={12566--12571},
  year={2024},
}

@misc{qwen3tts,
  title={Qwen3-TTS Technical Report},
  author={Hu, Hangrui and Zhu, Xinfa and He, Ting and Guo, Dake and Zhang, Bin and Wang, Xiong and Guo, Zhifang and Jiang, Ziyue and Hao, Hongkun and Guo, Zishan and others},
  journal={arXiv preprint arXiv:2601.15621},
  year={2026}
}

@inproceedings{flashattention,
  title={Flashattention: Fast and memory-efficient exact attention with io-awareness},
  author={Dao, Tri and Fu, Dan and Ermon, Stefano and Rudra, Atri and R{\'e}, Christopher},
  booktitle={Conference on Neural Information Processing Systems (NeurIPS)},
  year={2022}
}

@misc{prolific,
    author={Prolific},
    title = {The participants for this paper were recruited using {P}rolific},
    year={2014},
    url={https://www.prolific.com/},
    note = {{A}ccessed: 02.2026}
}

@misc{qwen3,
  title={Qwen3 technical report},
  author={Yang, An and Li, Anfeng and Yang, Baosong and Zhang, Beichen and Hui, Binyuan and Zheng, Bo and Yu, Bowen and Gao, Chang and Huang, Chengen and Lv, Chenxu and others},
  journal={arXiv preprint arXiv:2505.09388},
  year={2025}
}

@article{bpe,
  title={A new algorithm for data compression},
  author={Gage, Philip},
  journal={The C Users Journal},
  volume={12},
  number={2},
  pages={23--38},
  year={1994},
  publisher={R \& D Publications, Inc. Lawrence, KS, USA}
}

@article{seedtts,
  title={Seed-tts: A family of high-quality versatile speech generation models},
  author={Anastassiou, Philip and Chen, Jiawei and Chen, Jitong and Chen, Yuanzhe and Chen, Zhuo and Chen, Ziyi and Cong, Jian and Deng, Lelai and Ding, Chuang and Gao, Lu and others},
  journal={arXiv preprint arXiv:2406.02430},
  year={2024}
}

@inproceedings{hyperspherical,
  title={Hyperspherical latents improve continuous-token autoregressive generation},
  author={Ke, Guolin and Xue, Hui},
  booktitle={International conference on learning representations (ICLR)},
  year={2026}
}

\end{document}